\pdfoutput=1
\documentclass[aps,pra,10pt,twocolumn,showpacs,superscriptaddress,preprintnumbers]{revtex4-1}
\usepackage{amsmath}
\usepackage{latexsym}
\usepackage{amssymb}
\usepackage{bm}
\usepackage{graphics,epstopdf}
\usepackage{color}
\usepackage{hyperref}

\usepackage{newlfont}
\usepackage{amsfonts}
\usepackage{amsthm}
\usepackage{graphicx}
\usepackage{epsfig}

\renewcommand{\t}[1]{\textrm{#1}}%
\usepackage{times}%

\usepackage[up]{subfigure}

\newcommand{\be}{\begin{equation}}%
\newcommand{\ee}{\end{equation}}%
\newcommand{\bc}{\begin{center}}%
\newcommand{\ec}{\end{center}}%
\newcommand{\bea}{\begin{eqnarray}}%
\newcommand{\eea}{\end{eqnarray}}%
\newcommand{\ba}{\begin{array}}%
\newcommand{\ea}{\end{array}}%

\begin{document}

\title{Monogamy, polygamy, and other properties of entanglement of purification}
\author{Shrobona Bagchi }
\email{shrobona@hri.res.in}
\affiliation{Quantum Information and Computation Group, \\
Harish-Chandra Research Institute,\\
Chhatnag Road, Jhunsi, Allahabad-211019, India.}
\author{Arun Kumar Pati}
\email{akpati@hri.res.in}
\affiliation{Quantum Information and Computation Group, \\
Harish-Chandra Research Institute,\\
Chhatnag Road, Jhunsi, Allahabad-211019, India.}

\preprint{HRI-P-15-02-001} 

\date{\today}

\begin{abstract}
For bipartite pure and mixed quantum states, in addition to the quantum mutual information, there is another
measure of total correlation, namely, the entanglement of purification. We study the monogamy, polygamy, and
additivity properties of the entanglement of purification for pure and mixed states. In this paper, we show that,
in contrast to the quantum mutual information which is strictly monogamous for any tripartite pure states, the
entanglement of purification is polygamous for the same. This shows that there can be genuinely two types of
total correlation across any bipartite cross in a pure tripartite state. Furthermore, we find the lower bound and
actual values of the entanglement of purification for different classes of tripartite and higher-dimensional bipartite
mixed states. Thereafter, we show that if entanglement of purification is not additive on tensor product states,
it is actually subadditive. Using these results, we identify some states which are additive on tensor products for
entanglement of purification. The implications of these findings on the quantum advantage of dense coding are
briefly discussed, whereby we show that for tripartite pure states, it is strictly monogamous and if it is nonadditive,
then it is superadditive on tensor product states.
\end{abstract}

\maketitle

\section{ Introduction}
In quantum world there are different kinds of correlations between multiparticle quantum states. Understanding the nature of correlations is one of the
challenges in the development of quantum information
science. Given a bipartite or multipartite state one usually tries to characterize the amount of classical correlation, quantum correlation and quantum entanglement 
contained in the composite system. Different correlations can arise depending on the state preparation procedure and
measurements performed on the system. These correlations can account for many counter-intuitive
features in the quantum world.
In particular, entanglement is a physical property that has been successfully employed to interpret several phenomena which cannot be understood
using the laws of classical physics \cite{Horodecki1}. It has also been identified as the basic ingredient for different quantum communication protocols like super-dense coding
\cite{Wiesner}, quantum teleportation \cite{Brassard}, quantum cryptography \cite{Bennett}, remote-state preparation \cite{Pati, Bennett2}
and quantum computational tasks such as the one-way quantum computer \cite{bri}.


One fundamental property of quantum correlations in multiparty quantum states is that it can be monogamous \cite{Coffman}. To state this in a qualitative way, if a correlation 
measure is monogamous, then
this says that in a composite quantum state, if two subsystems are more correlated with each other, then they will share a less amount of correlation with the other subsystems 
with respect to that measure of correlation. In other words, it puts a restriction on the shareability of correlation between the different parties of a composite quantum state. 
Specifically, 
if the two subsystems are maximally quantum correlated with each other, then they cannot get correlated to any other subsystem at the same time.
The measures of classical correlation are never monogamous and  therefore are considered to be freely shareable. But, not all measures of quantum correlation
satisfy monogamy \cite{Osborne,Adesso,Hiroshima,Seevinck,Lee}. For example, the square of concurrence and the squashed entanglement satisfy the  monogamy inequality \cite{Matthias1},
whereas the relative entropy of entanglement, the entanglement of formation and other measures do not satisfy monogamy in general. Recently, it has been shown that the monogamous 
character is not an intrinsic property of other quantum correlation measures. In particular, the quantum discord \cite{Zurek} for tripartite states does not obey monogamy in general 
\cite{Prabhu,gio,alex}. However, interestingly, though a quantum correlation measure may not satisfy monogamy, yet the quantum correlation measure raised to a power will certainly 
obey monogamy \cite{Salini}. It has been shown that the square of the concurrence, which is a monotonic function of entanglement of formation, is monogamous. Similarly, it has been shown that 
the square of the quantum discord also satisfies monogamy.
The concept of monogamy, not only is important from fundamental point of view, it finds practical importance too.
For example, the monogamy of quantum correlations plays a crucial role in the security of quantum cryptography \cite{Gisin1}.

While the monogamy is an important property to study for various correlation measures, still there remains other desirable properties that the 
correlation measures are expected to obey from the perspective of being physically meaningful. 
One such property is the additivity on tensor product of density matrices \cite{Shor1}. 
The property of additivity on tensor product states dictates that a correlation measure is an additive measure if the value of that measure on the tensor product of density matrices is 
simply equal to the addition of the values of that correlation measure on the individual density matrices forming the tensor product state. 
The quantum mutual information is an additive measure of total correlation and the squashed entanglement is another additive measure of quantum
correlation \cite{Matthias}.
However, all correlation measures are not yet proved to be additive \cite{Werner1}. 
There are measures of entanglement and 
capacity of channels that have been proved to be non-additive \cite{Werner,Hastings,Hayden1,Smith}. For example the relative entropy of entanglement is proved to be non-additive 
\cite{Shor2} and there is
strong indication that the bipartite distillable entanglement is also non-additive \cite{Werner1}. Also, the additivity of entanglement of formation still remains an open question,
 and it is conjectured to obey a strong super-additivity condition \cite{Pomeransky}. Thus, the question of additivity of the different correlation measures is
 one of the intriguing and yet to be solved question in the realm of quantum information theory.


A measure of total correlation was proposed by
Terhal {\it et al}. \cite{Terhal} called the entanglement of purification. It should be emphasized that entanglement of purification
is not a measure of entanglement, but a measure of total correlation defined in units of pure state entanglement.  This definition was motivated operationally, 
trying to see if quantum states could be
constructed from EPR pairs, i.e. the Einstein-Podolsky-Rosen
pairs, with vanishing amount of communication asymptotically. It
is based on the entanglement-separability paradigm, 
trying to capture the classical and quantum correlations in an unified way. 
It was shown to be satisfying the properties of a genuine measure of total correlation. 
Also, a monogamy relation involving the entanglement of purification and the quantum advantage of dense coding was given by Horodecki {\it et al}. \cite{Horodecki}.
However, the conditions for the monogamy or polygamy nature of entanglement of purification have not been found yet. The present paper is motivated from the fact that the mutual 
information, 
a measure of total correlation is monogamous for any tripartite pure states \cite{gio}. Therefore, if the entanglement of purification is a measure of total correlation can it be 
strictly monogamous for all tripartite pure states? We find that the entanglement of purification of a tripartite pure state $\rho_{ABC}$ across $A:BC$ partition is never less than
its sum for the reduced density matrices $\rho_{AB}$ and $\rho_{AC}$, and is mostly polygamous. This observation calls for further investigation in understanding
the nature of correlation captured by the entanglement of purification. At first, we prove that similar to the mutual information, the entanglement of purification does not increase 
upon discarding ancilla. Thereafter, we explore the monogamy, polygamy and the additivity properties of the entanglement of purification for pure as well as mixed tripartite
states. Furthermore, we find analytically the lower bound and actual value of the entanglement of purification for different classes of mixed states.
We also present some conditions for the monogamy of entanglement of purification in terms of monogamy of entanglement of formation and other entropic inequalities.
We use these properties of entanglement of purification to explore the monogamy and additivity properties of the quantum advantage of dense coding. 
The above definition as a theory of 'all correlation' may have important applications in quantum information theory.

The paper is organized as follows. In section II, we provide the definition of the monogamy of correlations.
In section III, we discuss the measures of total correlation namely the quantum mutual information 
and the entanglement of purification, mentioning specifically the monogamy 
properties of the quantum mutual information. Here we also state the definition of interaction information and discuss some of its properties briefly.
Then, we move on to find the relation between the entanglement of purification and the quantum mutual information of the purified state in section IV.
Here, we discuss what happens to the entanglement of purification upon discarding of subsystem of a composite quantum system. Thereafter, we
obtain the lower bounds and exact values of entanglement of purification for some mixed quantum states, specifically for a class of tripartite states and
higher dimensional bipartite states. In section V,
we derive the results for the monogamy and polygamy nature  of the entanglement of purification for pure as well as mixed states, extending to the case of $n$ parties. 
Here, we discuss a relation between the monogamy of entanglement of purification with that of the entanglement of formation and quantum discord in
the case of tripartite pure states. Also, in this section, the monogamy conditions for the mixed states are explored  with specific examples
and cases where the the states are polygamous. 
In section VI, we find out that if entanglement of purification is not additive on tensor product of density matrices, then 
it has to be a sub-additive quantity. Next, using the results we derive in the previous sections,
we find out the monogamy, super-additivity (on tensor products) properties for the quantum advantage of dense coding in section VII, 
where we also obtain the upper bounds for some states and identify the states with no quantum advantage of dense coding. We end with conclusions and outlook in section VIII.

\section{Monogamy and Polygamy of correlations}
Monogamy is a property of a multiparticle quantum state that can be studied with respect to a particular correlation measure. 
It is an important property that tells us about the nature of the correlation at our disposal, in particular, whether it is freely shareable or not. 
Classical correlations \cite{Henderson} are always polygamous, whereas certain quantum correlation
measures satisfy this property and some others do not \cite{Adesso,Hiroshima,Prabhu,Matthias}. For example, the quantum discord is not in general a monogamous quantity for even 
some cases of the pure tripartite states, whereas the total correlation given by the quantum mutual information is strictly monogamous for all tripartite pure states.
Therefore, the monogamy or polygamy nature of the total correlation measure that supposedly contains some amount of quantum and classical
correlation is an important question to consider. 
Now, according to the definition of monogamy, it is a property which does not allow the free sharing of correlation between the 
subparts of a composite system. Mathematically, if a correlation measure $Q(\rho)$ satisfies 
\begin{equation}
 Q(A:BC)\geq Q(A:B)+Q(A:C)  
\end{equation}
for any tripartite state $\rho_{ABC}$, then the correlation measure is called monogamous, otherwise it is called polygamous. This definition can be extended to the case of 
$n$ parties as well. A correlation measure $Q$ is said to be $n$ partite monogamous if the following inequality is satisfied
\begin{equation}
  Q(A_1:A_2..A_n)\geq  Q(A_1:A_2)+Q(A_1:A_3)+..Q(A_1:A_n)\nonumber
\end{equation}
and otherwise it is called $n$ partite polygamous.

\section{ Measures of total correlation}

We consider multiparticle quantum system with each subsystem defined on a finite dimensional Hilbert space ${\cal H}$. Let ${\cal L}(\cal H)$ be the 
set of all linear operators acting on ${\cal H}$ and $D({\cal H})$ be the set of all density operators $\rho$ with $\rho \ge 0$ and $\t Tr(\rho) =1$. 
The composite state $\rho_{ABC} \in D({\cal H}_{ABC} )$ is a general state that may contain classical and quantum correlations including entanglement. 
The von-Neumann entropy for a density operator $\rho_A$ is defined as 
$S(A) = -{\t Tr}(\rho_A \log_2 \rho_A)$, where $ \rho_{A}=Tr_{BC}(\rho_{ABC})$.
In this section we discuss two important measures of total correlation in the bipartite scenario, namely, 
the quantum mutual information and the entanglement of purification. The measures of total correlation try to capture quantitatively the total correlations comprising of the 
classical as well as the quantum correlations in a bipartite state $\rho_{AB} = {\t Tr}_C (\rho_{ABC})$.

\subsection{ Quantum Mutual Information}

The quantum mutual information is a measure of total correlation in a quantum system.
It is a straightforward generalization of the classical mutual information. The quantum mutual information is obtained by just replacing the Shannon entropy 
by the von-Neumann entropy for the respective terms in the expression for the classical mutual information. 
Thus, for a bipartite quantum state, the quantum mutual information of the state $\rho_{AB}$ is defined as $ I(A:B)= S(A)-S(A|B)$, where $
S(A|B) = S(AB)- S(B)$ is the quantum conditional entropy \cite{Cerf}. Quantum mutual information satisfies some natural properties, all of which,  
a total correlation measure is expected to satisfy. 
First, it never increases 
upon discarding of quantum systems, i.e., $I(A:BC)\geq I(A:B)$. Secondly, the quantum mutual information is additive on tensor product of density matrices, which is 
$I(AC:BD)=I(A:B)+I(C:D)$ for $\rho_{AB}\otimes\sigma_{CD}$. 
Apart from these, the monogamy properties of the mutual information have been studied in Ref.\cite{gio}. There, it was shown that a necessary and sufficient condition 
for the monogamy of quantum mutual information can be stated in terms of the interaction information \cite{Prabhu,gio}. Specifically, it can be shown that for any pure tripartite state 
$\vert\Psi\rangle_{ABC}$, we have 
\begin{align}
I(A:B) + I(A:C) = I(A:BC), \nonumber
\end{align}
which implies that the quantum mutual information is strictly monogamous for a pure tripartite state. The necessary and sufficient criteria for quantum mutual information to be 
monogamous for mixed tripartite state is that the interaction information should be positive \cite{gio}. 
In classical information theory, interaction information of state $\rho_{ABC}$ is defined as $\tilde{I}(\rho_{ABC})= H(AB)+H(BC)+H(AC)-H(A)-H(B)-H(C)-H(ABC)$, where $H(AB)$ denote the Shannon entropies 
\cite{Thomas}.
Replacing the Shannon entropies by the von-Neumann entropies we obtain the quantum generalization of the interaction information. The quantum interaction information is therefore
nothing but $\tilde{I}(\rho_{ABC})= S(AB)+S(BC)+S(AC)-S(A)-S(B)-S(C)-S(ABC)$, where $S(AB)$ denote the von-Neumann entropy of the density matrix $\rho_{AB}$. 
Interaction information is a measure of the effect of 
the presence of a third party $C$ on the amount of correlations shared by the other two parties as it is given by the difference between the information shared between the parties 
$A$ and $B$ when $C$ is present and when $C$ is not present. Quantum interaction information can be positive as well as negative. It is invariant under the action of local unitaries 
and non-increasing under the action of unilocal measurements \cite{Prabhu}. It has been used to provide necessary and sufficient conditions for the monogamy of quantum discord in
Ref.\cite{Prabhu}.
Quantum mutual information is an important measure of correlation and finds application in
a large number of settings primarily in studying the channel capacities \cite{Bennett1,Benjamin}. Also, an operational interpretation has been given of the quantum mutual 
information in Ref.\cite{Groissman}. There, it was interpreted as the total amount of randomness or noise needed to erase the correlations in a bipartite quantum state 
completely.

\subsection{ Entanglement of purification}

The entanglement of purification is a measure of total correlation along a bipartition in a quantum state, \cite{Terhal} defined 
using the notion of the entanglement separability paradigm. Interestingly, in this approach the authors in \cite{Terhal} have treated both the quantum entanglement and the classical correlation in a unified framework, 
by defining a measure of total correlation namely the entanglement of purification in units of pure state entanglement.
By their definition, the entanglement of purification is expressed as the entanglement of the purified version of the mixed state as follows. Suppose we have a mixed state $\rho_{AB}$, and we purify it 
to a pure state $\vert\Psi\rangle_{ABA'B'}$. Then, the entanglement of purification is defined as
\begin{equation}
E_p(A:B)=\min_{A'B'} E_f(AA':BB'),
\end{equation}
where $E_p(A:B)$ denotes the entanglement of purification of the state $\rho_{AB}$ across $A:B$ partition, and $E_f(AA':BB')$ is the entanglement of 
formation across the bipartition $ AA':BB'$ of the pure state $\vert\Psi\rangle_{ABA'B'}$, 
obtained from $\rho_{AB}$ by any standard
purification procedure such as $\vert\Psi_s\rangle_{AA':BB'}= \sum_{i}\sqrt{\lambda_i}\vert \Psi_i\rangle_{AB}\otimes\vert 0\rangle_{A'}\vert i\rangle_{B'}$. Here, the $\lambda_i$  
are the Schmidt coefficients and $ \vert \Psi_i\rangle $ are the corresponding Schmidt vectors in $\cal H_{AB}$. 
The above expression can be reformulated in terms of the trace preserving completely positive (TCP) maps, 
since every quantum operation can be written in terms of the TPCP maps. Following Ref.\cite{Terhal}, from Eq(4), we 
get $E_p(A:B)$ of $\rho_{AB}$ as the following minimum over unitary matrices as 
\begin{align}
E_p(A:B)=\min_{U_{A'B'}}E_f(AA':BB'),
\end{align}
where $E_f(AA':BB')$ is the entanglement of formation across the $AA':BB'$ partition of the pure state 
${(I_{AB}\otimes U_{A'B'})(\vert\Psi_s\rangle\langle\Psi_s\vert)(I_{AB}\otimes U_{A'B'})^{\dagger}}$ obtained from $\rho_{AB}$ by a standard purification procedure and then acting 
unitary matrices over the ancilla part. This is nothing but the entropy $\min_{U_{A'B'}}S(Tr_{AA'}((I_{AB}\otimes U_{A'B'})(\vert\Psi_s\rangle\langle\Psi_s\vert)(I_{AB}\otimes U_{A'B'})^{\dagger}))$. Now by tracing out the
$ AA'$ part from the pure state as well as the unitary operator, one obtains the following equivalent form of entanglement of purification in terms of the TCP map
\begin{eqnarray}
E_p(A:B)=\min_{\Lambda_{B'}}S ((I_{B}\otimes\Lambda_{B'})(\mu_{BB'}(\rho_{AB})));\nonumber\\
\Lambda_{B'}(\nu) = {\t Tr}_{A'}(U_{A'B'}(\nu_{B'}\otimes\vert0\rangle\langle 0\vert_{A'})U^\dagger_{A'B'});\nonumber\\
\mu_{BB'}(\rho_{AB})= {\t Tr}_{AA'}(\vert\Psi\rangle\langle\Psi\vert),
\end{eqnarray}
where $ \Lambda_{B'}$ is a TCP map. The above form is derived in \cite{Terhal}. Therefore, the 
minimization over unitary matrices in Eq(3) is now represented as a minimization over all TPCP maps $\Lambda_{B'}$, since a TCP map is equivalently represented as an unitary transformation on the larger system followed by 
tracing over the ancilla.
It was shown that the above optimization can be successfully performed in a Hilbert space of a limited dimension $ d_{A'}=d_{AB}$ and $d_{B'}=d_{AB}^2 $, due to the result by 
Terhal {\it et al}. \cite{Terhal}.
For pure states, the entanglement of purification is equal to the entanglement of formation and for a mixed state $\rho_{AB}$, one has $E_p(A:B)\geq E_f(A:B)$.
Alongside, the authors have introduced the regularised entanglement of purification $E_p^\infty(A:B)$. It was shown that the asymptotic cost of preparing $n$ copies of 
$\rho_{AB}$ from singlets using only local operations and an asymptotically vanishing
amount of quantum or classical communication is equal to the regularised entanglement of purification.
This implies that the regularised entanglement of purification is actually the 
entanglement cost (with $LO_q$) of the quantum states $\rho$ on $\cal H_d\otimes\cal H_d$ \cite{Terhal}, i.e., $E_{LO_q}(A:B)=E_p^\infty(A:B)$.
Later, from an operational point of view it was shown that if it is additive on tensor product states then $ E^\infty_P(A:B)$ is actually the optimal visible compression rate for mixed 
states \cite{Hayashi}. Other operational interpretations have been explored for this quantity. In particular, the regularized entanglement of purification was shown to be equal to the
entanglement assisted noisy channel capacity \cite{Nilanjana}. On another note it was shown that the regularized entanglement of purification $E_{LO_q}(A:B)$ gives the communication cost 
of simulating a channel without the presence of prior entanglement \cite{Shor}. However, the entanglement of purification is mostly an unexplored quantity since 
it is a difficult quantity to calculate analytically owing to the optimization needed to be done in a larger Hilbert space. But, using the monogamy property of entanglement,
the authors in Ref.\cite{Matthias1} have found the entanglement of purification for a class of bipartite states supported in symmetric or antisymmetric subspaces analytically to be 
$S(A)$. However, one of the unanswered question regarding the entanglement of purification is the property of additivity. 
It is still not known whether the entanglement of purification is additive on tensor product states or not. But, some progress has been made in this direction by, where
entanglement of purification has been proved to be non-additive within a certain numerical tolerance \cite{Chen}. The entanglement of purification has been related to some other 
information theoretic quantities as well. It has also been shown that the entanglement of purification is related to the partial quantum information, through its monogamy relation 
with the quantum advantage of dense coding \cite{Horodecki}.

\section{ Entanglement of purification in terms of quantum mutual information
: Lower bound and exact values}

The entanglement of purification can be rewritten in terms of the quantum mutual information.
For the pure state $\vert\Psi\rangle_{ABA'B'}$, which is the
optimally purified state for the mixed state $\rho_{AB}$ for evaluating the entanglement of purification, the quantum mutual information between parties $AA'$ and $BB'$ is given by 
$ I(AA':BB')= S(AA')+S(BB')-S(AA'BB')$. Since $\vert\Psi\rangle_{ABA'B'}$ is a pure state, we have
\begin{equation}
 E_p(A:B) = \frac{I(AA':BB')}{2}.\nonumber
\end{equation}
Therefore, the entanglement of purification is actually half of the optimised quantum mutual information of the purified version of the mixed density
matrix. The above equations are then used to prove a better lower bound for the entanglement of purification. Before that, we prove an important property of entanglement of 
purification, an attribute of a measure of total correlation.
\vskip 10pt
{\textbf{ Proposition 1}}: The entanglement of purification never increases upon discarding of quantum system, i.e.,
\begin{equation}
 E_p(A:BC)\geq E_p(A:B).
\end{equation}

\textit{Proof}: 
If $\rho_{ABC}$ is pure, then $E_p(A:BC)=S(A)$.  Also, we know that $E_p(A:B)\leq S(A)$. This leads to $E_p(A:BC)\geq E_p(A:B)$.
In case of mixed states $\rho_{ABC}$, we note that the set of all the pure states for calculating $E_p(A:BC)$ is a subset of the set of all pure states taken for calculating
$E_p(A:B)$.
This clearly implies that $ min[I(AA':BB')]\leq min[I(AA':BC(BC)')]$. From here we thus conclude that $E_p(A:BC)\geq E_p(A:B)$. Thus, like
the quantum mutual information, the entanglement of purification also never increases upon discarding of quantum systems. This is a desired property that the total
correlation should not increase upon discarding of quantum system. It is easily seen that the equality condition holds when $\rho_{AB}$ is supported in the symmetric
or antisymmetric subspace.

We now state some simple inequalities for entanglement of purification which will be later used for deriving the monogamy and polygamy conditions for it. Let 
$\vert\Psi\rangle_{ABA'B'}$ be the optimal pure state for evaluating the entanglement of purification of $\rho_{AB}$. 
Using the sub-additivity of conditional entropy \cite{Chuang} for a composite quantum system of four parties, i.e., 
$S(AB\vert A'B')\leq S(A\vert A')+S(B\vert B')$, we get $S(ABA'B')-S(AB)\leq S(AA')-S(A)+S(BB')-S(B)$. But we know $E_p(A:B)=S(AA')=S(BB')$ and $S(ABA'B')=0$, 
since according to the definition of entanglement of purification $\rho_{ABA'B'}$ is a pure state. Using this in the above inequality,
we get $2E_p(A:B)\geq I(A:B)$. Therefore, we have the following lower bound on $E_p(A:B)$
\begin{equation}
 E_p(A:B)\geq \frac{I(A:B)}{2}.
\end{equation}
Extending this to the asymptotic limit, one easily obtains $E_p^{\infty}(A:B)=E_{LO_q}(A:B)\geq\frac{I(A:B)}{2}$, by using the fact that the quantum mutual information is additive on 
tensor product of quantum states. The above lower bound was known for the entanglement of purification, but only in the asymptotic limit,
and it was obtained from an operational point of view in \cite{Terhal}. Here we obtain this bound for a single copy of $\rho_{AB}$, 
and easily extend this to the asymptotic limit as $E_{LO_q}(A:B)\geq \frac{I(A:B)}{2}$ and get back the result given in Ref.\cite{Terhal}. 
Also, the lower bound given in \cite{Terhal} for a single copy of $\rho_{AB}$ is $E_f(A:B)$. However, we know that for some states one has 
$E_f(A:B)\leq \frac{I({A:B})}{2}$. Therefore, for these states we get a better lower bound
for a single copy of $\rho_{AB}$. Now, we use the equation for entanglement of purification in terms of 
quantum mutual information to derive a lower bound for tripartite mixed states which is different from half of its quantum mutual information.
\vskip 10pt
{\textbf {Proposition 2}}:
For any pure or mixed tripartite quantum state:
\begin{equation}
 E_p(A:BC)\geq S(A)-\frac{1}{2}[S(A\vert B)+S(A\vert C)].
\end{equation}

{\textit {Proof}}:
Let $\vert\Psi\rangle_{ABCA'D'}$ be the optimal pure state for evaluating the entanglement of purification of $\rho_{ABC}$. 
Therefore, we have $E_p(A:BC)= \frac{I(AA':BCD')}{2}$.
Note that the quantum mutual information of pure states satisfy the monogamy equality condition. Therefore, $E_p(A:BC)= \frac{I(AA':B)}{2}+\frac{I(AA':CD')}{2}$. 
Again, the mutual information is non-increasing upon discarding of quantum systems, hence we have 
\begin{align}
E_p(A:BC)\geq \frac{I(A:B)}{2}+\frac{I(A:C)}{2}.
\end{align} 
This implies
$E_p(A:BC)\geq S(A)- (\frac{S(A\vert B)}{2}+\frac{S(A\vert C)}{2})$. In general, from the previous literature we know that $E_p(A:BC)\geq \frac{I(A:BC)}{2} $.
However, for the states with $I(A:BC)\leq I(A:B)+I(A:C)$, i.e, with the negative interaction information, we then have
$E_p(A:BC)\geq \frac{I(A:B)}{2}+\frac{I(A:C)}{2}\geq\frac{I(A:BC)}{2}$. Therefore, for these class of states, the entanglement of purification is upper and lower bounded as
$S(A)\geq E_p(A:BC)\geq S(A)- (\frac{S(A\vert B)}{2}+\frac{S(A\vert C)}{2})$. 
Extending this to the asymptotic limit we obtain $E_{LO_q}(A:BC)\geq \frac{I(A:B)}{2}+\frac{I(A:C)}{2}$, using the fact that quantum mutual information is 
additive on tensor product of density matrices.
We note that the tripartite quantum states with negative interaction information are always 
polygamous for the quantum mutual information. Therefore, for these states, the above bound is always greater than the previous bound $\frac{I(A:BC)}{2}$. This may give
a better lower bound than $\frac{I(A:B)}{2}$ or the regularised classical mutual information \cite{Terhal} for states consisting of quantum as well as classical correlations, 
depending on the negativity of interaction information. One may extend this to the case of $n$ parties as well, such that for a $n$ partite density matrices $\rho_{A_1A_2...A_n}$,
we get $E_p(A_1:A_2A_3..A_n)\geq \max[\frac{(I(A_1:A_iA_j..)}{2}+\frac{(I(A_1:A_kA_l..)}{2}]$ etc. where one takes all possible combinations of bipartitions between $A_1A_2...A_n$
(keeping the node $A_1$ same for the reduced density matrices)
to achieve the maximum value of the lower bound. Therefore, the quantum states with negative interaction information across any bipartition will have either the regularised 
classical mutual information or this as the better lower bound than half of its quantum mutual information.
\vskip 10pt
{\textbf {Corollary}}:
The entanglement of purification for the class of tripartite mixed states satisfying the sub-additivity equality condition is given by $S(A)$.
\vskip 10pt
{\textit {Proof}}: From the previous paragraph we see that when $S(A\vert B)+S(A\vert C)=0$, we get $E_p(A:BC)\geq S(A)$. But again, from the upper bound of
entanglement of purification we have $E_p(A:BC)\leq S(A) $. Therefore combining the above two equations, one obtains $E_p(A:B)=S(A)$ for the states which
satisfy the strong sub-additivity equality condition. Also, we know that mixtures
of the tripartite mixed states each satisfying the strong sub-additivity equality condition and satisfying an additional constraint of biorthogonality if the third party is
traced out, satisfy the strong sub-additivity equality, and hence their entanglement of purification is also $S(A)$. Hence the proof. 
The structure of the states obeying the sub-additivity equality condition has been precisely given in Ref.\cite{Hayden}. There it was shown that every separable state can 
be extended to a state that obeys the sub-additivity equality condition. Therefore, from these observations
we can comment that all separable states can be extended to a tripartite mixed state which has the maximum amount of total correlation as $S(A)$.
From the viewpoint of the structure of the states \cite{Hayden},
the structure states satisfying the SSA equality has been given as $ \rho_{ABC} = \bigoplus_j q_j\rho_{Ab_j^L}\otimes\rho_{b_j^RC}$,
with states $\rho_{Ab_j^L} $ on Hilbert space $ H_A\otimes H_{b^L_j}$ and $\rho_{b^R_jC}$ on $ H_{b^R_j}\otimes H_C$ with probability distribution $q_j$. 
Thus, all states of this form and all extensions of this class of states
have the maximal amount of total correlation given by the entanglement of purification as $S(A)$.
Now we discuss the lower bound and exact values with some specific examples as given below.

\begin{figure}
\includegraphics[scale=0.63]{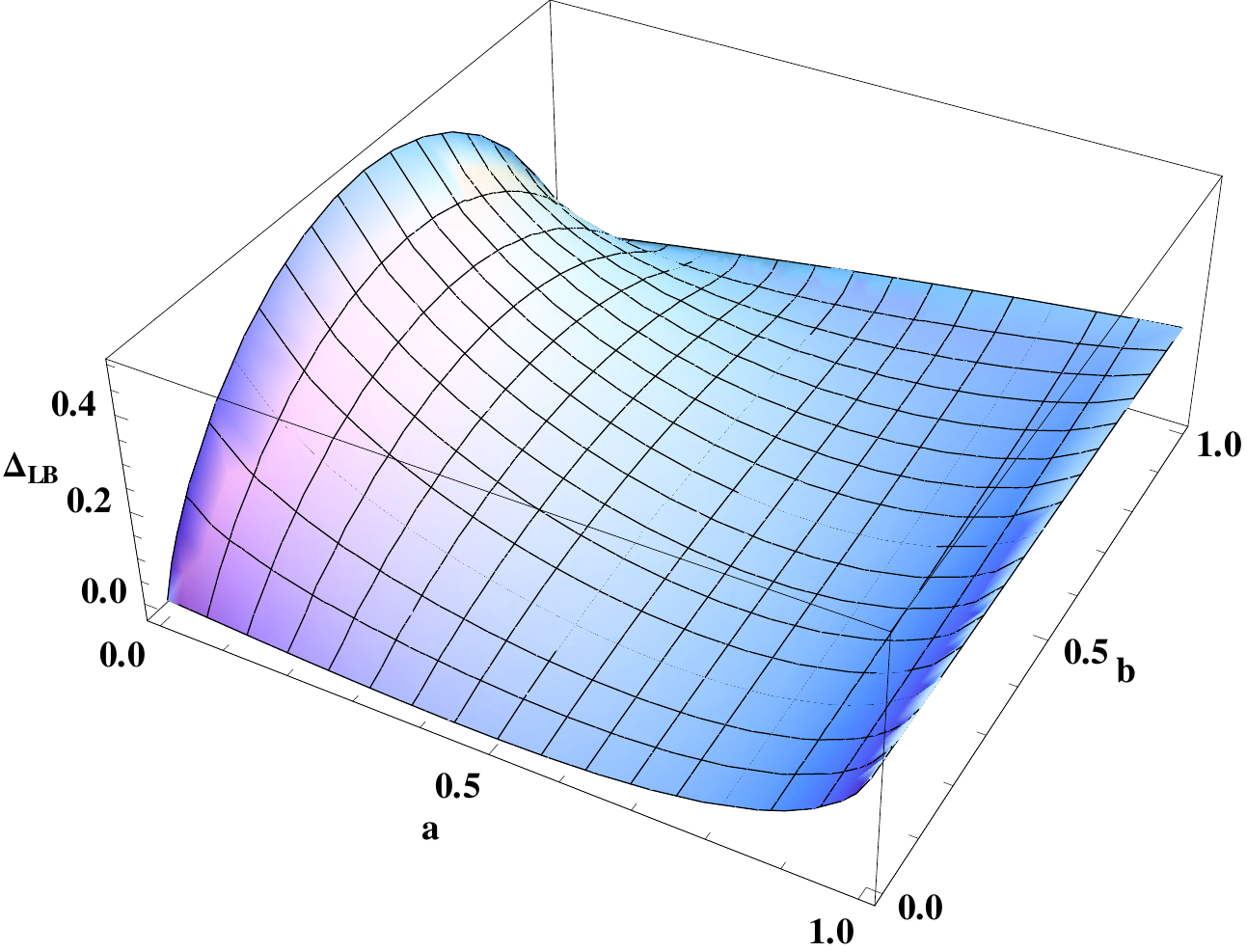}
\caption{Difference between lower bounds for state $ p\vert W\rangle\langle W\vert +(1-p)[a\vert 000\rangle\langle 000\vert +(1-a)\vert 111\rangle \langle 111\vert]$.
The difference between the new lower bound and the previous one is always positive in this case.}
\end{figure}

\textit {Examples of exact values}:

First we state the value of entanglement of purification for the following class of bipartite mixed states. 
The entanglement of purification of the states satisfying the Araki-Lieb equality condition is $S(A)$.
We know $S(A)\geq E_p(A:B)\geq \frac{1}{2}I(A:B)$. But $\frac{1}{2}I(A:B)=S(A)+\frac{1}{2}[S(B)-S(A)-S(AB)]$. The states satisfying the Araki-Lieb equality
condition have $S(B)-S(A)=S(AB)$. Then, we have $S(A)\geq E_p(A:B)\geq S(A)$. Therefore, $ E_p(A:B)= S(A)$ for these states.
The structure of states satisfying the Araki-Lieb equality condition is given in Ref.\cite{Zhang}. There, it was shown that the states satisfy the Araki-Lieb equality condition
if and only if the following conditions are satisfied. First, $\cal{H_A}$ can be factorized as $\cal {H_L}\otimes\cal{H_R}$ and secondly 
$\rho_{AB}= \rho_L\otimes\vert\Psi_{RB}\rangle\langle\Psi_{RB}\vert$, where $\vert\Psi_{RB}\rangle\in\cal{H_R}\otimes\cal{H_B}$. 
The structure of such states that satisfy the Araki-Lieb equality condition is therefore of the form $\rho_{AB}= \rho_L\otimes\vert\Psi_{RB}\rangle\langle\Psi_{RB}\vert$.
Therefore, the value of entanglement of purification for these states is $S(A)$.

For the case of tripartite states, the entanglement of purification of states of the form 
$\rho_{ABC}= p\vert GHZ\rangle\langle GHZ\vert^{\underline{+}}+(1-p)[b\vert 000\rangle\langle 000\vert+(1-b)\vert 111\rangle \langle 111\vert]$
is $S(A)$ for all values of $\{p,a,b\} \in [0,1]$, where $\vert GHZ\rangle^{\underline{+}} = \sqrt{a}\vert 000\rangle \underline{+} \sqrt{(1-a)}\vert111\rangle$ is the 
generalized GHZ state \cite{GHZ}. This holds for $n$ party as well, i.e., for the following state
$\rho_{ABC}= p\vert GHZ_n\rangle\langle GHZ_n\vert^{\underline{+}} +(1-p)[b\vert 0\rangle\langle 0\vert^{\otimes n} +(1-b)\vert 1\rangle \langle 1\vert^{\otimes n}]$
where $ \vert GHZ_n\rangle^{\underline{+}} =\sqrt{a}\vert 0\rangle^{\otimes n} \underline{+} \sqrt{(1-a)}\vert1\rangle^{\otimes n}$.
The proof is as follows. We know that for tripartite states $E_p(A:BC)\geq \frac{1}{2}[I(A:B)+I(A:C)]$. For the state given above, $I(A:B)+I(A:C)=2I(A:B)=2[S(A)+S(B)-S(AB)]=2S(A)$.
The first equality follows owing to the symmetry of the state between parties $B$ and $C$. The third equality follows from the fact that the nonzero eigenvalues of the density 
matrices $\rho_{AB}$ and $\rho_{B}$ are exactly equal. Therefore, for the given state $ S(A)\geq E_p(A:BC)\geq S(A)$. Thus, $E_p(A:BC)=S(A)$. Let us consider another 
example. The tripartite mixed state as a mixture of the  $\vert GHZ\rangle^+$ and $\vert GHZ\rangle^-$, i.e., if 
$\rho_{ABC}= p\vert GHZ\rangle\langle GHZ\vert^+ +(1-p)\vert GHZ\rangle\langle GHZ\vert^-$ then it also has $E_p(A:B)= S(A)$ according to
our previous argument. Here, the states are generalized $\vert GHZ\rangle$ states. And similar to the above,
this is also true for the arbitrary mixture of $n$ partite generalized $\vert GHZ\rangle$ states.

\textit {Examples of lower bounds}:
Among other examples, for the tripartite states of the form
$ \rho_{ABC}=p\vert W\rangle\langle W\vert +(1-p)[a\vert 000\rangle\langle 000\vert +(1-a)\vert 111\rangle \langle 111\vert]$,
where $\vert W\rangle=\frac{1}{\sqrt{3}}[\vert 100\rangle+\vert 010\rangle+\vert 001\rangle]$ is the $\vert W\rangle$ state, a
better lower bound
is provided by $\frac{1}{2}[I(A:B)+I(A:C)]\geq \frac{1}{2}I(A:BC)$, since the quantum mutual information is polygamous for these classes of states. This holds even for the regularised
version of the entanglement of purification, i.e., $E_{LO_q}(A:BC)\geq \frac{1}{2}[I(A:B)+I(A:C)]$, owing to the additivity of the 
quantum mutual information on tensor product of density matrices. The difference $\Delta_{LB}$ between the two lower bounds equal to
$\frac{1}{2}[I(A:B)+I(A:C)-I(A:BC)]$ is plotted in Fig 1, which shows that it is always positive. Again we may consider the state
$ \rho_{ABC}= p\vert W\rangle\langle W\vert +\frac{(1-p)}{8}I_3$ and the difference between the lower bounds are plotted in Fig 2.

\begin{figure}
\includegraphics[scale=0.83]{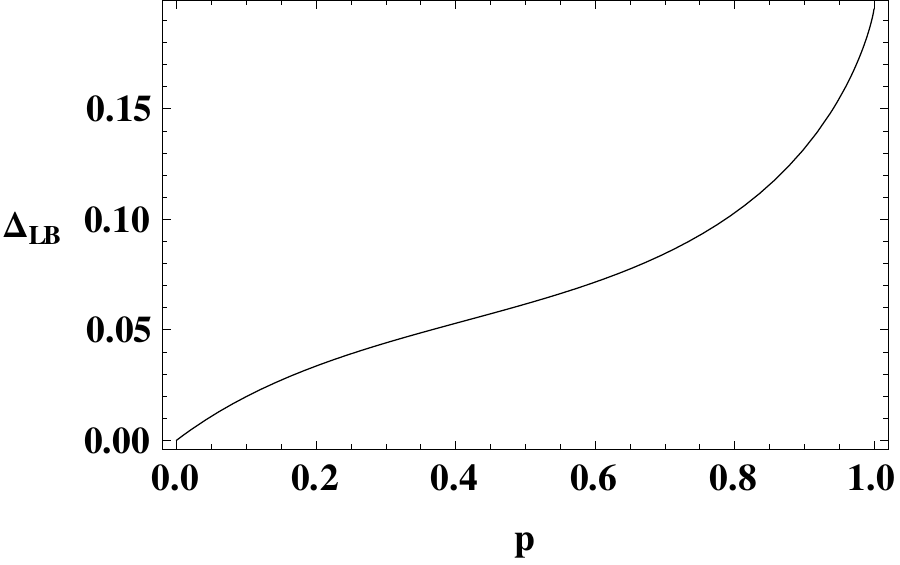}
\caption{Difference between lower bounds for state $ \rho= p\vert W\rangle\langle W\vert +\frac{(1-p)}{8}I_3$. The difference between the new and the old lower bound is always positive
here. The difference in lower bounds is given by the amount of polygamy of quantum mutual information.}
\end{figure}

One can use the polygamy of the quantum mutual information to lower bound the entanglement of purification in higher dimensional bipartite states. If a sub-party is of higher 
dimension, and if the quantum mutual information is polygamous for the lower dimensional subparts obtained by breaking the higher dimensional subparty, then it gives  a better
lower bound for the entanglement of purification than just half of the quantum mutual information of the state $\rho_{AB}$. 

Suppose for a $ 2^n$ dimensional party $B$ in $\rho_{AB}$, we break it down into two lower dimensional subparties $B_1$ and $B_2$ \cite{Cornello}. Then, from Eq(8) we have
$E_p(A:B)\geq \frac{1}{2}[I(A:B_1)+I(A:B_2)]$.
For negative interaction information between $B_1$ and $B_2$, i.e., $S(AB_1)+S(AB_2)+S(B_1B_2)-S(A)-S(B_1)-S(B_2)-S(AB_1B_2)< 0$, 
the R.H.S is greater than $\frac{I(A:B)}{2}$ \cite{gio}. Thus it gives a better lower bound. We may say 
that this better lower bound arises as a result of a second order polygamy relation of quantum mutual information. 
One can easily extend to the asymptotic limit as well, thus we obtain the lower bound  
$E_{LO_q}(A:B)\geq \frac{1}{2}[I(A:B_1)+I(A:B_2)]> \frac{I(A:B)}{2}$. For these states $ E_{LO_q}(A:B)$ quantifies more correlation than $\frac{I(A:B)}{2}$ as given in the original paper.
For these states, one now has to compare the quantity $\frac{1}{2}[I(A:B_1)+I(A:B_2)]$ with the classical mutual information for obtaining a better lower bound.
The above equation can also be written as 
\begin{equation}
E_p(A:B)\geq S(A)-\frac{1}{2}[S(A\vert B_1)+S(A\vert B_2)]. \nonumber
\end{equation} 
From this equation we can say that for the $2^n$ dimensional party $B$ 
in the bipartite state $\rho_{AB}$, if the internal structure of $B$ is such that across any subpartition inside it, the sub-additivity equality condition is satisfied then 
the entanglement of purification of that state is $S(A)$. Therefore, with the aid of the new lower bound as half of the summation of the quantum mutual information of the subparties, 
we are able to conclude about the new exact values of entanglement of purification for these classes of the higher dimensional bipartite states.

\section{Monogamy and polygamy of entanglement of purification}

Here we explore various conditions under which the entanglement of purification will be polygamous or monogamous for pure and mixed states.

\subsection{ Monogamy and polygamy of entanglement of purification for pure tripartite states}
\vskip 10pt
{\textbf {Theorem 1}}:
The entanglement of purification is polygamous for a tripartite pure state $\rho_{ABC}$:
\begin{equation}
E_p(A:B)+E_p(A:C)\geq E_p(A: BC).
\end{equation}
\textit {Proof}:
From Eq(6) we know that $E_p(A:B)\geq \frac{I(A:B)}{2}$. Therefore, we have
$E_p(A:B)+E_p(A:C)\geq \frac{I(A:B)}{2}+\frac{I(A:C)}{2}$. In
case of the tripartite pure state $\rho_{ABC}$ the right hand side of the inequality just gives $S(A)$. This implies that
\begin{equation}
 E_p(A:B)+E_p(A:C)\geq S(A).\nonumber
\end{equation}
Since for pure tripartite state $\rho_{ABC}$, $E_p(A: BC)=S(A)$, we obtain:
\begin{equation}
 E_p(A:B)+E_p(A:C)\geq E_p(A:BC).\nonumber
\end{equation}
This shows the polygamous nature of the entanglement of purification for pure tripartite state $\rho_{ABC}$.
One can directly see that the same relation holds for the regularised entanglement of purification:
$E_{LO_{q}}(A:B)+E_{LO_{q}}(A:C)\geq E_{LO_{q}}(A:BC)$, i.e., the regularised entanglement of purification is also a polygamous quantity.
This proves that entanglement of purification for any tripartite pure state is in general a polygamous quantity.
An implication of this is that the sum of the asymptotic entanglement cost of preparing $\rho_{AB}$ and $\rho_{AC}$ will 
not be restricted by the asymptotic cost of preparing $\rho_{A:BC}$.

The polygamy inequality above shows that there can be states satisfying the equality condition in the inequality. To analyse the states that may satisfy the equality condition we 
find a following relation to the monogamy of entanglement of formation for those states.
Given a pure state $\rho_{ABC}$, if entanglement of formation violates monogamy, then entanglement of purification will
violate monogamy equality for the same. However the converse is not true.
The proof is as follows. If entanglement of formation $E_f(A:BC)$ violates monogamy for some pure state $\rho_{ABC}$, then we have
$E_f(A:BC)< E_f(A:B)+E_f(A:C)$.
But for a pure state $\rho_{ABC}$, we know that $E_f(A:BC)=E_p(A:BC)$. Therefore, replacing this
in the above equation we get $ E_p(A:BC)< E_f(A:B)+E_f(A:C)$. Also, it is known that for 
any state $\rho_{AB}$, we have $E_f(A:B)\leq E_p(A:B) $.
This implies $E_p(A:BC)< E_p(A:B)+E_p(A:C)$
which shows that the entanglement of purification also violates monogamy. Hence the proof. However the vice versa may not be true.
We know that for pure states the monogamy of entanglement of formation is equivalent to
the monogamy of quantum discord \cite{gio}. Therefore, we conclude that the polygamy of quantum discord will also imply the polygamy of entanglement of purification likewise.
In other words, monogamy of entanglement of formation or quantum discord is a necessary condition for the tripartite state $\rho_{ABC}$ to satisfy the 
monogamy equality condition for entanglement of purification.
Now let us try to compare the monogamy inequality of the entanglement of formation with the entanglement of purification for
mixed tripartite state $\rho_{ABC}$. Before that, we define a quantity called correlation of classical and quantum origin $E_{cq}(A:B)$ of the state $\rho_{AB}$ as
\begin{align}
E_{cq}(A:B) = E_{p}(A:B) - E_{f}(A:B).\nonumber
\end{align}
This quantity is positive for mixed states and vanishes for pure bipartite states. Intuitively, this may contain some classical 
correlation and some amount of quantum correlation beyond entanglement that is captured by the entanglement of formation.
From the definition it is clear that for a 
given mixed state $\rho_{ABC}$, if $E_{cq}(A:B)$ and $E_{f}(A:B)$ are monogamous (polygamous), then the entanglement of purification will be (monogamous) polygamous.
One can also show that for three-qubit states if the the correlation of classical and quantum origin obeys monogamy and entanglement of formation satisfies \cite{fanch} 
 \begin{align}
E_{f}(A:B) + E_{f}(A:C) \le 1.18\nonumber
\end{align}
then the entanglement of purification will obey a weak monogamy relation as given by
\begin{align}
E_{p}(A:B) + E_{p}(A:C) \le   E_{p}(A:BC) + 1.18.
\end{align}

\subsection{Mixed states}

The entanglement of purification $E_p(A:BC)$ of a mixed tripartite state $\rho_{ABC}$ is $\frac{I(AA':BC(BC)')}{2}$, where the optimal pure state of $\rho_{ABC}$ is
$\vert\Psi_{ABCA'(BC)'}\rangle$. Similarly, the entanglement of purification $E_p(A:B)$ of $\rho_{AB}$ is $\frac{I(AA'':BB'')}{2}$, where the optimal pure state for $\rho_{AB}$ is
$\vert\Phi_{ABA''B''}\rangle$,  and the entanglement of purification $E_p(A:C)$ of $\rho_{AC}$ is $\frac{I(AA''':CC''')}{2}$, where the optimal pure state for $\rho_{AC}$ is
$\vert\xi_{ACA'''C'''}\rangle$. Therefore, the monogamy inequality for a mixed tripartite state $\rho_{ABC}$ is
$I(AA':BC(BC)')\geq I(AA'':BB'')+I(AA''':CC''')$. But owing to the largely difficult optimization needed, we may not be able to check this equation directly. Instead, we analyze 
some specific cases of mixed states that are polygamous for entanglement of purification as follows.

At first we note that the tripartite mixed states satisfying the strong sub-additivity equality condition are polygamous for entanglement of purification. To see this,
let $\vert\Psi_{ABA'B'}\rangle$ and $\vert\Psi_{ACA''C''}\rangle$ be the optimal pure states for $\rho_{AB}$ and $\rho_{AC} $ respectively.
Then $E_p(A:B)+E_p(A:C)\geq \frac{1}{2}[I(A:B)+I(A:C) + I(AA':B')+I(AA'':C'')]$. But $ I(A:B)+I(A:C) = 2S(A)-(S(A\vert B)+S(A\vert C))$,
and if the strong sub-additivity equality condition is satisfied then we have  $S(A\vert B)+S(A\vert C) = 0$. 
Putting these in the equation, we get $ E_p(A:B)+E_p(A:C)\geq S(A) + \frac{1}{2}[I(AA':B')+I(AA'':C'')] $.
But the last two terms on the R.H.S are positive in general, as the quantum mutual information is always positive and vanishes only for the maximally mixed state. 
Also, we know $E_p(A:BC)= S(A)$. Thus, combining these inequalities together we obtain 
$ E_p(A:B)+E_p(A:C)\geq E_p(A:BC)$. Thus, the entanglement of purification is polygamous for the class of states that satisfy the strong sub-additivity equality.
Among other classes of states, if anyone of the reduced density matrices $\rho_{AB}$, $\rho_{AC}$  of a mixed state $\rho_{ABC}$ are entirely supported on
the symmetric or antisymmetric subspaces, then the state will violate monogamy of entanglement of purification.
This follows from the result by Winter \textit{et al}.\cite{Matthias1}. The entanglement of purification of such bipartite density matrices (with the same dimension for both parties) is 
$S(A)$. But the entanglement of purification of the tripartite mixed state is also $S(A)$ and in general $E_p(A:C)\geq 0$. Therefore, the polygamy inequality follows directly by 
combining the above observations. Also, any tripartite extension of bipartite mixed states that satisfy the Araki-Lieb equality condition for their von-Neumann entropy
is polygamous for entanglement of purification. We know that the states that satisfy the Araki Lieb equality condition have $E_p(A:B)= S(A)[S(B)]$. 
However, the other reduced density matrix has some non zero correlation and therefore non-zero entanglement of purification. Thus, in this case we have 
$E_p(A:B)+E_p(A:C)\geq S(A)=E_p(A:BC)$, making entanglement of purification a polygamous measure of total correlation. Though for pure tripartite states we could prove the general 
polygamy inequality, for mixed states it is not clear whether such general inequality exists or not.

Next, we discuss the relation to polygamy of quantum mutual information. 
Suppose $E_p(A:B)+E_p(A:C)= S(B\vert A)+S(C\vert A)+I(A:B)+I(A:C)$. This is greater than $ S(BC\vert A)+I(A:B)+I(A:C)$ which is again
greater than $ E_p(A:BC)+I(A:B)+I(A:C)-I(A:BC)$. From the above equations one can see that if the mutual information is polygamous,
then here the entanglement of purification becomes polygamous. Again, a sufficient condition for monogamy of $E_p$
is  $\frac{I(A:BC)}{2}\geq E_p(A:B)+E_p(A:C)$. This implies $\frac{I(A:BC)}{2}\geq \frac{I(A:B)}{2}+\frac{I(A:C)}{2}$, which is nothing but
$I(A:BC)\geq I(A:B)+I(A:C)$, i.e., the monogamy inequality for the quantum mutual information. This says that the states satisfying this particular sufficient
condition for monogamy of $E_p$ will also satisfy the monogamy inequality of quantum mutual information. 

\subsection{Polygamy of entanglement of purification for multiparty}

Now we investigate the polygamy of entanglement of purification in case of $n$ partite density matrices. The conditions for the polygamy for mixed states
also get translated here as sufficient conditions for polygamy. To put it in other words, the $n$-partite density matrices, pure or mixed, are polygamous if
any one of the reduced density matrices of the subsystem satisfy the Araki-Lieb equality condition, strong sub-additivity equality condition or is supported on the symmetric
or antisymmetric subspace. Now we state a simple sufficient condition for the polygamy of entanglement of purification and construct some examples.
\vskip 10pt
{\textbf{Proposition 3}}: All the $n$-partite states, pure or mixed with $\sum_{i=1}^n I(A:A_i)\geq 2S(A)$ are polygamous for entanglement of purification.
\vskip 10pt

\textit{Proof}: We have  $\sum_{i=1}^n E_p(A:A_i)\geq\frac{1}{2}[\sum_{i=1}^n I(A:A_i)]$. From this we get
$\sum_{i=1}^n E_p(A:A_i)\geq S(A)+\frac{1}{2}[\sum_{i=1}^n I(A:A_i)]-2S(A)$. Thus, we get the condition in the proposition as the sufficient condition for polygamy of
entanglement of purification. A large number of states will satisfy this condition, and thus will be polygamous. However, some states will violate this condition,
and it will be inconclusive about the polygamous nature in case of those states. 

Using the above relation, we easily see that
the $n$-party generalized $\vert GHZ\rangle $ and the $n$-party $ \vert W\rangle$ states are polygamous with respect to the entanglement of purification. We can explicitly see the
proofs as follows. We have the generalized $GHZ$ state as $\vert GHZ\rangle = \sqrt{p}\vert 0\rangle^{\otimes n} + \sqrt{1-p}\vert 1\rangle^{\otimes n}$, where $0\leq p\leq1$ 
\cite{GHZ}.
But we have obtained before that for tripartite pure states, $ E_p(A:A_1)+E_p(A:A_2)\geq S(A)$. Thus, it holds true for the tripartite generalized $\vert GHZ\rangle$ states as well.
Now for $n\geq 3$, we see that all the reduced density matrices are exactly the same and L.H.S becomes $\sum_{i=1}^n E_p(A:A_i)$. 
This is nothing but $ E_p(A:A_1)+E_p(A:A_2)+\sum_{i=3}^n E_p(A:A_i)$. Since each of the two party reduced density matrices are exactly the same as the two party reduced density matrices 
in the case of tripartite pure state, therefore using the above two equations we obtain $\sum_{i=1}^n E_p(A:A_i)\geq S(A)+\sum_{i=3}^n E_p(A:A_i)$. The last term on R.H.S is 
always positive. Therefore we obtain $ \sum_{i=1}^n E_p(A:A_i)\geq S(A)$, rendering the entanglement of purification polygamous for all $n$ in the case of generalized 
$\vert GHZ\rangle$ state. This is expected since every reduced density matrices share only classical correlation with the other reduced density matrices.
We now consider $\vert W\rangle = \frac{1}{\sqrt{n}}[\vert 10..0\rangle + \vert 01..0\rangle + .. ]$, where there are $n$ terms within the parenthesis \cite{Dur}.
We show that this state is also polygamous for all values of $n$. To see this, first we note that all the two party reduced density matrices $\rho_{AA_i}$ of this state 
are exactly same due to the symmetry of the state. Specifically each $\rho_{AA_i} = \frac{1}{n}[(n-2)\vert 00\rangle\langle 00\vert]+2\vert\Phi^+\rangle\langle\Phi^+\vert]$, where
$\vert\Phi^+\rangle=\frac{1}{\sqrt{2}}[\vert 10\rangle+\vert 01\rangle]$ is the Bell state. Now we calculate $\frac{1}{2}[\sum_{i=1}^n I(A:A_i)]=\frac{n}{2}I(A:A_1)$, since 
all the two party reduced density matrices are same. Evaluating the eigenvalues in terms of $n$, we find that $S(A)=S(A_1)=2\log_2 n-\log_2(n-1)$ and $S(AA_1)=2\log_2 n-1-\log_2(n-2)$.
Putting these values in the equation above, we get
$\frac{n}{2}I(A:A_1)-S(A)=\frac{n}{2}+\frac{n}{2}\log_2 (n-2) +(n-1)\log_2 \frac{n}{n-1}$. This value is always positive for all values of $n>2$. 
Thus combining the earlier result of tripartite pure state with the above finding, we conclude that the entanglement of purification is polygamous 
for $n$ party $\vert W\rangle$ state. 

Likewise the case for mixed states, where we state some conditions relating monogamy of entanglement of purification with that of quantum mutual information,
we now state a proposition connecting the polygamy of quantum mutual information to the polygamy of entanglement of purification for a pure state of $n$ parties.
\vskip 10pt
{\textbf{Proposition 4}}: All the $n$ party pure states for which the quantum mutual information is $(n-1)$ partite polygamous for at least any one of the $(n-1)$ party reduced density 
matrices of the pure state, is $n$ partite polygamous for both the entanglement of purification as well as the quantum mutual information.
\vskip 10pt
{\textit{Proof}}: Note that for $n$ partite pure state, we have $\sum_{i=2}^n E_p(A_1:A_i)\geq \frac{1}{2}\sum_{i=2}^n I(A_1:A_i)$. Now, let us take a reduced density
matrix $\rho_{A_1A_2...A_{n-1}}$ to be polygamous for quantum mutual information, i.e., $\sum_{i=2}^{n-1} I(A_1:A_i)\geq I(A_1:A_2...A_{n-1})$. Then, we have 
$I(A_1:A_n)+\sum_{i=2}^{n-1} I(A_1:A_i)\geq I(A_1:A_n)+I(A_1:A_2...A_{n-1})$. Since the $n$ partite quantum state we are considering is a pure state,
therefore by virtue of monogamy of quantum mutual information, the R.H.S. of this equation is nothing but $ I(A_1:A_2A_3...A_n)$. But, we know for a pure state 
$ I(A_1:A_2A_3...A_n)=2S(A_1)$. From here it then follows that $\sum_{i=2}^n E_p(A_1:A_i)\geq S(A_1)$ and also $\sum_{i=2}^n I(A_1:A_i)\geq 2S(A_1)$. 
These two equations are just the equations of polygamy for the entanglement of purification and the quantum mutual respectively for a $n$ partite pure state. It is easy to see that
one could take any one of the possible $(n-1)$ different reduced density matrices possible of the $n$ partite pure state (keeping the node $A_1$ intact for each reduced density matrix)
as the one polygamous for the quantum mutual information and eventually get back the polygamy equation for both the entanglement of purification and quantum mutual information.
As a specific example of this proposition, we easily see that all the four party pure states with negative interaction information across any two pair of its bipartite reduced density 
matrices, are polygamous for entanglement of purification.

\section{ Sub-additivity on tensor products}
Additivity is a desirable property to hold for a given measure of total correlation. Quantum mutual information is an additive measure of correlation, however entanglement of 
purification may not be an additive measure. Using strong numerical support this has been shown in Ref.\cite{Chen}. Here we prove that if it is non-additive then it has to be a
sub-additive quantity. We have the following theorem.
\vskip 5pt
{\textbf {Theorem 2}}:
The entanglement of purification is sub-additive in the tensor product of density matrices, i.e., for a tensor product density matrix ${\rho_{AB}\otimes\sigma_{CD}}$, the following equation
holds
\begin{equation}
 E_p(AC: BD)\leq E_p(A: B)+E_p(C: D).\nonumber
\end{equation}
with equality if and only if the optimal pure state for the tensor product of density matrices is
the tensor product of optimal pure states of the corresponding density matrices upto a local unitary equivalence.
\vskip 10pt
{\textit {Proof}}:
Let us suppose $\vert\Psi_{ABA'B'}\rangle$ and $\vert\Phi_{CDC'D'}\rangle$ are the optimal purification for $\rho_{AB}$ and $\sigma_{CD}$ 
corresponding to the value of entanglement of 
purification. Then $ \vert\Psi_{ABA'B'}\rangle\otimes\vert\Phi_{CDC'D'}\rangle $ is a valid purification for $\rho_{AB}\otimes\sigma_{CD} $, 
however not generally the optimal one. Now, we know 
that $ E_p(A: B)=\frac{I(AA': BB')}{2}$ and $ E_p(C: D)=\frac{I(CC': DD')}{2}$ . 
Adding these two quantities we get $E_p(A: B)+E_p(C: D)=\frac{I(AA': BB')}{2}+\frac{I(CC': DD')}{2}$.
But the quantum mutual information is additive on tensor product of quantum states. Therefore, $\frac{I(AA': BB')}{2}+\frac{I(CC': DD')}{2}=\frac{I(AA'CC': BB'DD')}{2}$ 
where $I(AA'CC': BB'DD')$ is the quantum mutual information of the state $\vert\Psi_{ABA'B'}\rangle\otimes\vert\Phi_{CDC'D'}\rangle$. Thus, we have
\begin{align} \nonumber E_p(A: B)+E_p(C: D) =\frac{I(AA'CC': BB'DD')}{2}.\nonumber
\end{align}
Since $\vert\Psi_{ABA'B'}\rangle\otimes\vert\Phi_{CDC'D'}\rangle$ is only one such purification of $\rho_{AB}\otimes\sigma_{CD}$ 
and the optimization for $E_p(AC:BD)$ is over all possible purifications of $\rho_{AB}\otimes\sigma_{CD}$ denoted by the set of pure states 
$\{\vert\xi_{ABCDA''B''}\rangle\}$, therefore we have
\begin{align}\nonumber
 \min_{A''B''} \frac{I(ACA'': BDB'')}{2}
 \leq \frac{I(ACA'C': BDB'D')}{2}, \nonumber
\end{align}
where $I(ACA'': BDB'')$ is the quantum mutual information of any such purification $ \vert\xi_{ABCDA''B''}\rangle\ $ and the minimum is over all such purification of 
$\rho_{AB}\otimes\sigma_{CD}$ by the addition of ancilla part $A''B''$ to it. Hence we easily see that the above equation is nothing but the following inequality,
\begin{align}\nonumber
E_p(AC: BD) \nonumber
\leq \frac{I(ACA'C':BDB'D')}{2},\nonumber
\end{align}
which directly implies that, $E_p(AC: BD)\leq E_p(A:B)+E_p(C:D)$ for the four partite tensor product density matrix $\rho_{AB}\otimes\sigma_{CD}$.
Now, in the following paragraph we check the equality condition. 

While checking the equality condition, we now omit the subscripts and write $\vert\Psi_{ABA'B'}\rangle $ 
as $\vert\Psi\rangle$, $\vert\Phi_{CDC'D'}\rangle$ as $ \vert\Phi\rangle$ and $\vert\xi_{ABCDA''B''}\rangle$ as $\vert\xi\rangle$ for simplicity.
First, we check that if $ \vert\xi\rangle=\vert\Psi\rangle\otimes\vert\Phi\rangle$, then whether the 
dimensionality of the optimal purifying state agrees with the dimension  of the Hilbert space of the ancilla part, as given in Ref.\cite {Terhal}. 
We note that if $ \vert\xi\rangle=\vert\Psi\rangle\otimes\vert\Phi\rangle$, then $d_{A''}(\vert\xi\rangle)=d_{A'}(\vert\Psi\rangle)d_{C'}(\vert\Phi\rangle),
d_{B''}(\vert\xi\rangle)=d_{B'}(\vert\Psi\rangle)d_{D'}(\vert\Phi\rangle)$.
According to the theorem given in Ref.\cite {Terhal}, $d_{A'}(\vert\Psi\rangle)=d_{AB}(\rho_{AB})$, $d_{C'}(\vert\Phi\rangle)= d_{CD}(\sigma_{CD})$ and 
$ d_{A''}(\vert\xi\rangle)=d_{ABCD}(\rho_{AB}\otimes\sigma_{CD})$. Similarly by the same theorem, we have 
$d_{B'}(\vert\Psi\rangle)=d_{AB}^2(\rho_{AB})$, $d_{D'}(\vert\Phi\rangle)= d_{CD}^2(\sigma_{CD})$ and 
$ d_{B''}(\vert\xi\rangle)=d_{ABCD}^2(\rho_{AB}\otimes\sigma_{CD})$. Now, we verify if the above two equations are consistent with dimensions proposed in Ref.\cite{Terhal}
for $\vert\xi\rangle$. Putting the values of $d_{A'}$ and $d_{B'}$ in terms of $d_{AB}$, we get 
$d_{A''}(\vert\xi\rangle)=d_{AB}(\rho_{AB})d_{CD}(\sigma_{CD})$ and $d_{B''}(\vert\xi\rangle)=d_{AB}^2(\rho_{AB})d_{CD}^2(\sigma_{CD})$. These values can be reframed as
the dimensions of the tensor product of the corresponding density matrices, i.e., $d_{AB}(\rho_{AB})d_{CD}(\sigma_{CD})= d_{ABCD}(\rho_{AB}\otimes\sigma_{CD})$. Similarly 
$d_{AB}^2(\rho_{AB})d_{CD}^2(\sigma_{CD})= d_{ABCD}^2(\rho_{AB}\otimes\sigma_{CD})$. This holds true even when 
$\vert\xi\rangle=U_{A'C'}\otimes U_{B'D'}\vert\Psi\rangle\otimes\vert\Phi\rangle$, since the unitary matrices do not map density matrices from Hilbert space of a given dimension
to that of a different dimension. 
This shows that the dimensions are in agreement with those given by the theorem in 
Ref.\cite{Terhal}.

We now move on to the equality condition for the mutual information. 
For this purpose, let us note that if $\vert\xi\rangle=U_{A'C'}\otimes U_{B'D'}\vert\Psi\rangle\otimes\vert\Phi\rangle$, 
then owing to the additivity of quantum mutual information and its invariance under the action of local unitaries, 
one has $I(ACA'':BDB'')=I(AA':BB')+I(CC'':DD')$, where the mutual information terms are that of $\vert\xi\rangle$, $\vert\Psi\rangle$, 
and $\vert\Phi\rangle$ respectively.
This implies that $E_p(AC:BD)=E_p(A:B)+E_p(C:D)$ for $\rho_{AB}\otimes\sigma_{CD}$. This proves the if part of 
theorem above. 

For the only if condition we see that if $ \vert\xi\rangle\neq U_{A'C'}\otimes U_{B'D'}\vert\Psi\rangle\otimes\vert\Phi\rangle$, 
then $I(\vert\xi\rangle)\neq I(\vert\Psi\rangle\otimes\vert\Phi\rangle)$. This is because, the action of a non-local unitary will 
change the probability distributions of the reduced density matrices and thus will change the value of quantum mutual information across $ACA'C':BDB'D'$ partition. As a result, 
the equality holds only if the optimal pure state for the tensor product of the density matrices is the tensor 
product of corresponding optimal pure states, upto the local unitary equivalence $U_{A'C'}\otimes U_{B'D'}$. 
Therefore, we see that if the entanglement of purification is non-additive, it is actually sub-additive. Thus, the above theorem rules out the 
super-additivity of entanglement of purification. The sub-additivity has been shown numerically in Ref.\cite{Chen} for the Werner states.
It is important to note that, according to the result by authors in Ref.\cite{Terhal}, one is guaranteed to find the optimal pure state in the Hilbert space of the aforementioned 
dimensionality. In that case our equality condition holds for the tensor product of the optimal pure states. However it does not rule out the existence of optimal pure states in 
Hilbert space of other dimensions. Thus, in addition to the optimal pure state in Hilbert space of the dimensions given by the theorem, one may find other optimal pure states in 
Hilbert space of higher or lower dimension. In particular, one might be able to find optimal pure state in Hilbert space of lower dimension.
As an example we have the Werner state and its optimal pure state for entanglement of purification can be found in Hilbert space of dimensions $4\times 4$ as proved numerically
in Ref.\cite{Terhal}.

Using the results we have obtained on entanglement of purification, 
we identify the classes of states that are additive on tensor products for the entanglement of purification as follows. We see that
the bipartite states satisfying the equality condition in Araki-Lieb inequality, the higher dimensional bipartite states satisfying the equality condition in strong sub-additivity
when any party of it can be broken down into two lower dimensional subparties, the tripartite states satisfying the strong sub-additivity equality condition
are additive on tensor products for entanglement of purification.
Thus, for the above class of states, the regularised entanglement of purification and their optimal visible
compression rate is given by the entanglement of purification. Apart from this, we are able to also draw the conclusion that
the entanglement of purification is additive on tensor products if and only if it is also super-additive on tensor products for all quantum states.
However, whether there can be states $\rho_{AB}\otimes\sigma_{CD}$ for which $ E_p(AC:BD) < E_p(A:B)_{\rho_{AB}}+E_p(C:D)_{\sigma_{CD}}$ is still an open question. We note that the 
question of non-additivity is now reduced to only the sub-additivity condition, ruling out the possibility of
$E_p(AC:BD) > E_p(A:B)_{\rho_{AB}}+E_p(C:D)_{\sigma_{CD}}$ for $\rho_{AB}\otimes\sigma_{CD}$.

\section{ Implications on the quantum advantage of dense coding}

Quantum dense coding is a quantum communication protocol where one sends classical information beyond the classical capacity of the
quantum channel with the help of a quantum state shared between two distant observers, and a noiseless quantum channel. The quantum advantage of dense coding is the
increase in the rate of classical information transmission due to shared entanglement. Mathematically, the quantum advantage of dense coding of a quantum state $\rho_{AB}$is 
defined in terms of the
coherent information as $\Delta(A\rangle B) = S(B)-inf_{\Lambda_A}S[(\Lambda_A\otimes I_B)\rho_{AB}]=sup_{\Lambda_A}I'(A\rangle B)$, 
where the infimum or supremum is performed over all the maps $\Lambda_A$ acting on the state $\rho_{AB}$ and $ I'(A\rangle B)=S(B)-S(AB)$ is the coherent information of $\rho_{AB}$.
There, it was proved that the quantum advantage of dense coding is a non-negative quantity. Again, 
a quantum state is said to be dense codeable if the above quantity $ \Delta(A\rangle B)$ is strictly positive.
 It was shown in the paper by Horodecki \cite{Horodecki} that it suffices to consider
only the extremal TPCP maps in evaluating the infimum or supremum for the above quantity, owing to the concavity of the von-Neumann entropy. It was also shown that the
quantum advantage of dense coding may be non-additive, though not proved definitely. Apart from the aforementioned properties, the quantum advantage of dense coding
was shown to obey a monogamy relation with the entanglement of purification as $ S(B)\geq \Delta(A\rangle B)+E_p(B:C)$ \cite{Horodecki}, for any tripartite state $\rho_{ABC}$, with equality
for pure tripartite states.

Therefore, from the monogamy inequality and the polygamy of
entanglement of purification for pure tripartite states as well as some of the mixed tripartite states mentioned here previously, it follows that
\begin{equation}
\triangle(B\rangle A)+\triangle(C\rangle A)\leq \triangle(BC\rangle A),\nonumber
\end{equation}
implying that the quantum advantage of dense coding is strictly monogamous for the tripartite pure states as well as the other tripartite mixed states mentioned previously. 
This property is straight forwardly carried over to the asymptotic limit as well.
Thus, we have $\triangle^{\infty}(B\rangle A)+\triangle^{\infty}(C\rangle A)\leq \triangle^{\infty}(BC\rangle A)$ for those same set of states. Also, it is easy to see that for the 
mixed states satisfying SSA equality 
condition, the symmetric (antisymmetric) subspace condition and the states satisfying the Araki-Lieb equality condition and the cases for the $n$ partite pure states, monogamy is 
followed.

In the same way as that of the entanglement of purification, we conclude that the quantum advantage of dense coding is super-additive on tensor product of density matrices, i.e., 
for a four partite tensor product state $\rho_{AB}\otimes\sigma_{CD}$, we have the following equation
\begin{equation}
 \Delta(AC\rangle BD)\geq \Delta(A\rangle B) +\Delta(C\rangle D).\nonumber
\end{equation}
The proof is as follows. By definition, we have $\Delta(A>B)=sup_{\Lambda_A}I'(A>B)$. Thus, for the density matrix $\rho_{AB}\otimes\sigma_{CD}$ we have 
$\Delta(A\rangle B)+\Delta(C\rangle D)=sup_{\Lambda_A}I'(A\rangle B)+sup_{\Lambda_C}I'(C\rangle D)= 
sup_{\Lambda_{A}\otimes\Lambda_{C}}I'(AC\rangle BD)$.
The second equation follows from the fact that the von-Neumann entropies are additive on tensor products of density matrices. Again for $\rho_{AB}\otimes\sigma_{CD}$, by definition we have 
$\Delta(AC\rangle BD)=sup_{\Lambda_{AC}}I'(AC\rangle BD)$.
However, the optimization for $\rho_{AB}\otimes\rho_{CD}$ is over all $\Lambda_{AC}$, and $\{\Lambda_A\otimes\Lambda_C\}$ is only a subset of $\{\Lambda_{AC}\}$. Thus, 
$sup_{\Lambda_{AC}}I'(AC\rangle BD\geq sup_{\Lambda_A\otimes\Lambda_C}I'(AC\rangle BD)$ for the same four partite product state $\rho_{AB}\otimes\sigma_{CD}$.
With the last equation we arrive at the super-additivity equation
for the quantum advantage of dense coding for tensor product states of the form $\rho_{AB}\otimes\sigma_{CD}$, 
i.e., $\Delta(AC\rangle BD)\geq \Delta(A\rangle B) +\Delta(C\rangle D)$ for $\rho_{AB}\otimes\sigma_{CD}$.

Not only super-additivity, but the monogamy inequality with entanglement of purification has other implications on the quantum advantage of dense coding as well. 
From the lower bound and some of the actual value of entanglement of purification, 
using the property of monogamy with it and non-negativity of the quantum advantage of dense coding, we can identify some of the quantum states that have no 
quantum advantage of dense and also put an upper bound on it for some specific cases.

Let $\rho_{ABCD}$ be a quantum state, such that the sub-additivity equality condition is satisfied for the reduced density matrix $\rho_{ABC}$, i.e., $S(B\vert A)+S(B\vert C)=0$.
Then, from the monogamy inequality with entanglement of purification, we get $ S(B)\geq \Delta(D\rangle B)+E_p(B:AC)$. But, in this case $E_p(B:AC)=S(B)$. Thus, putting this
value, we have $ \Delta(D\rangle B)\leq 0$. But, since $ \Delta(D\rangle B)\geq 0$, thus we have $ \Delta(D\rangle B)=0$ for the states $\rho_{BD}$, i.e., the quantum advantage of
dense coding vanishes precisely for these states. Similarly, for any tripartite state, pure or mixed $\rho_{ABC}$, if the state $\rho_{BC}$ satisfies the Araki-Lieb equality condition,
then the quantum advantage of dense coding $\Delta(A\rangle B)$ of $\rho_{AB}$ also becomes zero. Apart from the above exact values,
the lower bound on entanglement of purification puts an upper bound on the quantum advantage of dense coding via its monogamy relation with the quantum advantage of dense coding.

\section{ Conclusions and Outlook}
In this paper, we find that the monogamous nature of correlations is not unique to quantum correlations, but can also be the case for the total correlations for certain quantum 
states. Thus, monogamy is not a property of the quantum correlation alone. Contrary to the monogamy nature of the mutual 
information for tripartite pure states, we have proved that the entanglement of purification can be polygamous for such states.
This shows that even though the mutual information and the entanglement of purification are supposed to capture total correlation, the nature of these correlations
can be completely opposite at least for tripartite systems.
In case of pure and mixed states, the monogamy of entanglement of purification is related to the monogamy of entanglement of formation. Also, we have found a
necessary condition for monogamy of entanglement of purification for a special class of mixed states, in terms of the interaction information or the polygamy of the
quantum mutual information. A new lower bound of the entanglement of purification has been given for the tripartite mixed states and higher dimensional bipartite systems. 
Using the formula for the lower bound we have been able to find the exact values
of entanglement of purification for some classes of states. Furthermore,
in this paper we have also shown that if entanglement of purification is not additive, it has to be a sub-additive quantity. 
Using these results we have also shown that the quantum advantage of dense coding
is strictly monogamous for all tripartite pure states and it is super-additive on tensor products. We have also identified some of the quantum states with no 
quantum advantage of dense coding.
We have brought forward these important aspects of the measure of total correlation as well as that of the quantum advantage of dense coding to the forefront. 
These will help us understand better the nature of total and quantum correlations of composite quantum states. 
This calls for more explorations and a deeper understanding of the total correlation present in a composite
mixed state. The total correlation quantified by the mutual information can be split into quantum correlation and classical correlation. However,
we still do not know whether we can express the entanglement of purification as the sum of quantum and classical correlations. In view of the
polygamy nature of entanglement of purification, can it be the case that the entanglement of purification contains more classical like
correlation than the quantum correlation. This will be a topic of future investigation.

\section{ Acknowldgements}
SB acknowledges discussions with Ujjwal Sen, Aditi Sen de, Andreas Winter, Debbie Leung and Atri Bhattacharya. SB acknowledges cluster computing facility at HRI.
SB and AKP acknowledge financial support from DAE, Govt. of India.

\end{document}